\PassOptionsToPackage{unicode}{hyperref}
\PassOptionsToPackage{hyphens}{url}
\PassOptionsToPackage{dvipsnames,svgnames,x11names}{xcolor}
\documentclass[
  12pt,
]{article}
\usepackage{amsmath,amssymb}
\usepackage{iftex}
\ifPDFTeX
  \usepackage[T1]{fontenc}
  \usepackage[utf8]{inputenc}
  \usepackage{textcomp} 
\else 
  \usepackage{unicode-math} 
  \defaultfontfeatures{Scale=MatchLowercase}
  \defaultfontfeatures[\rmfamily]{Ligatures=TeX,Scale=1}
\fi
\usepackage{lmodern}
\ifPDFTeX\else
\fi
\IfFileExists{upquote.sty}{\usepackage{upquote}}{}
\IfFileExists{microtype.sty}{
  \usepackage[]{microtype}
  \UseMicrotypeSet[protrusion]{basicmath} 
}{}
\usepackage{xcolor}
\usepackage[margin=1in]{geometry}
\usepackage{graphicx}
\makeatletter
\def\maxwidth{\ifdim\Gin@nat@width>\linewidth\linewidth\else\Gin@nat@width\fi}
\def\maxheight{\ifdim\Gin@nat@height>\textheight\textheight\else\Gin@nat@height\fi}
\makeatother
\setkeys{Gin}{width=\maxwidth,height=\maxheight,keepaspectratio}
\makeatletter
\def\fps@figure{htbp}
\makeatother
\NewDocumentCommand\citeproctext{}{}
\NewDocumentCommand\citeproc{mm}{%
  \begingroup\def\citeproctext{#2}\cite{#1}\endgroup}
\makeatletter
 \let\@cite@ofmt\@firstofone
 \def\@biblabel#1{}
 \def\@cite#1#2{{#1\if@tempswa , #2\fi}}
\makeatother
\newlength{\cslhangindent}
\newlength{\csllabelwidth}
\newenvironment{CSLReferences}[2] 
 {\begin{list}{}{%
  \setlength{\itemindent}{0pt}
  \setlength{\leftmargin}{0pt}
  \setlength{\parsep}{0pt}
  \ifodd #1
   \setlength{\leftmargin}{\cslhangindent}
   \setlength{\itemindent}{-1\cslhangindent}
  \fi
  \setlength{\itemsep}{#2\baselineskip}}}
 {\end{list}}
\usepackage{calc}

\usepackage{amsmath}
\usepackage{amsthm}
\usepackage{amsmath}
\usepackage{amssymb}
\usepackage{bm}
\usepackage{booktabs}
\usepackage{braket}
\usepackage{caption}
\usepackage{float}
\usepackage[bottom,flushmargin,hang,multiple]{footmisc}
\usepackage{graphicx}
\usepackage{hyperref}
\usepackage{libertine}
\usepackage[libertine]{newtxmath}
\usepackage[mathcal]{euscript}
\usepackage{multirow}
\usepackage{quoting}
\usepackage{setspace}
\usepackage{soul}
\usepackage{subcaption}
\usepackage{tabularx}
\usepackage{titlesec}
\usepackage[utf8]{inputenc}
\ifLuaTeX
  \usepackage{selnolig}  
\fi
\usepackage{bookmark}
\IfFileExists{xurl.sty}{\usepackage{xurl}}{} 
\hypersetup{
  pdftitle={Stochastic, Dynamic, Fluid Autonomy in Agentic AI: Implications for Authorship, Inventorship, and Liability},
  colorlinks=true,
  linkcolor={Maroon},
  filecolor={Maroon},
  citecolor={Blue},
  urlcolor={blue},
  pdfcreator={LaTeX via pandoc}}

\title{Stochastic, Dynamic, Fluid Autonomy in Agentic AI: Implications
for Authorship, Inventorship, and Liability}
\author{}
\date{\vspace{-2.5em}}

\begin{document}
\maketitle

\vspace*{\fill}

\begin{center}
Anirban Mukherjee

Hannah Hanwen Chang

\bigskip

3 April, 2025
\end{center}

\vspace*{\fill}

\noindent \hrulefill

\noindent Anirban Mukherjee
(\href{mailto:anirban@avyayamholdings.com}{\nolinkurl{anirban@avyayamholdings.com}})
is Principal at Avyayam Holdings. Hannah H. Chang
(\href{mailto:hannahchang@smu.edu.sg}{\nolinkurl{hannahchang@smu.edu.sg}};
corresponding author) is Associate Professor of Marketing at the Lee
Kong Chian School of Business, Singapore Management University. This
research was supported by the Ministry of Education (MOE), Singapore,
under its Academic Research Fund (AcRF) Tier 2 Grant,
No.~MOE-T2EP40221-0008.

\newpage

\begin{center} 
\noindent \textbf{Abstract}
\end{center}

\noindent Agentic Artificial Intelligence (AI) systems, exemplified by
OpenAI's DeepResearch, autonomously pursue goals, adapting strategies
through implicit learning. Unlike traditional generative AI, which is
reactive to user prompts, agentic AI proactively orchestrates complex
workflows. It exhibits stochastic, dynamic, and fluid autonomy: its
steps and outputs vary probabilistically (stochastic), it evolves based
on prior interactions (dynamic), and it operates with significant
independence within human-defined parameters, adapting to context
(fluid). While this fosters complex, co-evolutionary human-machine
interactions capable of generating uniquely synthesized creative
outputs, it also irrevocably blurs boundaries---human and machine
contributions become irreducibly entangled in intertwined creative
processes. Consequently, agentic AI poses significant challenges to
legal frameworks reliant on clear attribution: authorship doctrines
struggle to disentangle ownership, intellectual property regimes strain
to accommodate recursively blended novelty, and liability models falter
as accountability diffuses across shifting loci of control. The central
issue is not the legal treatment of human versus machine contributions,
but the fundamental unmappability---the practical impossibility in many
cases---of accurately attributing specific creative elements to either
source. When retroactively parsing contributions becomes infeasible,
applying distinct standards based on origin becomes impracticable.
Therefore, we argue, legal and policy frameworks may need to treat human
and machine contributions as functionally equivalent---not for moral or
economic reasons, but as a pragmatic necessity.

\begin{center}\rule{0.5\linewidth}{0.5pt}\end{center}

\noindent Keywords: Agentic Artificial Intelligence, Autonomy, Machine
Creativity, Authorship, Copyright, Inventorship, Patent, Liability,
Tort.

\smallskip

\newpage
\doublespacing

\section{Introduction}\label{introduction}

Agentic Artificial Intelligence (AI) refers to AI systems capable of
autonomously\footnote{Within this analysis, \emph{agency} denotes a
  system's capacity to initiate goal-directed actions---whether through
  programmed imperatives or learned behaviors---while \emph{autonomy}
  refers to the degree of independence from direct human control in
  executing those actions. This distinction builds on established AI
  literature: Stan Franklin and Art Graesser,
  \citeproc{ref-franklin1996agent}{{``Is It an Agent, or Just a
  Program?: A Taxonomy for Autonomous Agents,''} \emph{International
  workshop on agent theories, architectures, and languages} (Springer
  1996)} define an agent as `a system situated within and a part of an
  environment that senses that environment and acts on it, over time, in
  pursuit of its agenda,' emphasizing goal-directed behavior (p.~25).
  Michael Wooldridge and Nicholas R Jennings,
  \citeproc{ref-wooldridge1995intelligent}{{``Intelligent Agents: Theory
  and Practice''} (1995) 10 The knowledge engineering review 115}
  distinguish explicitly between an agent's `pro-activeness'---its
  ability `to exhibit goal-directed behaviour by taking the
  initiative'---and its `autonomy,' defined as operating `without the
  direct intervention of humans' (pp.~3--4). Jenay M Beer, Arthur D Fisk
  and Wendy A Rogers, \citeproc{ref-beer2014toward}{{``Toward a
  Framework for Levels of Robot Autonomy in Human-Robot Interaction''}
  (2014) 3 Journal of human-robot interaction 74} characterize autonomy
  similarly as `the extent to which a system can carry out its own
  processes and operations without external control' (p.~77). Finally,
  Jeffrey M Bradshaw and others, \citeproc{ref-bradshaw2013seven}{{``The
  Seven Deadly Myths of" Autonomous Systems"''} (2013) 28 IEEE
  Intelligent Systems 54} highlight autonomy's multifaceted nature,
  distinguishing `self-sufficiency---the capability of an entity to take
  care of itself' from `self-directedness, or freedom from outside
  control,' further clarifying the conceptual space in which modern
  agentic AI operates (p.~2).} pursuing long-term goals, making
decisions, and executing complex workflows without continuous human
intervention. Although agentic AI shares conceptual roots with earlier
intelligent agents---goal-oriented software designed to sense and act
within an environment---\footnote{\citeproc{ref-wooldridge1995intelligent}{Wooldridge
  and Jennings (n 1)}.}and autonomous agents in multi-agent
systems,\footnote{\citeproc{ref-jennings1998roadmap}{Nicholas R
  Jennings, Katia Sycara and Michael Wooldridge, {``A Roadmap of Agent
  Research and Development''} (1998) 1 Autonomous agents and multi-agent
  systems 7}; \citeproc{ref-stone2000multiagent}{Peter Stone and Manuela
  Veloso, {``Multiagent Systems: A Survey from a Machine Learning
  Perspective''} (2000) 8 Autonomous Robots 345};
  \citeproc{ref-wooldridge2009introduction}{Michael Wooldridge, \emph{An
  Introduction to Multiagent Systems} (John wiley \& sons 2009)}.} it
represents a significant advancement. Historically, intelligent agents
were typically constrained to narrowly defined tasks, operating under
rigid rules and requiring constant human oversight.\footnote{\citeproc{ref-caballar2024aiagents}{Rina
  Diane Caballar, {``What Are {AI} Agents?''}
  \textless{}\url{https://spectrum.ieee.org/ai-agents}\textgreater{}};
  \citeproc{ref-clarke2019regulatory}{Roger Clarke, {``Regulatory
  Alternatives for AI''} (2019) 35 Computer Law \& Security Review 398}.}
In contrast, modern agentic AI systems leverage advanced technologies
such as reinforcement learning (RL), large language models (LLMs), and
sophisticated planning algorithms to interpret context, dynamically
adapt strategies, and proactively orchestrate multi-step
processes.\footnote{\citeproc{ref-shavit2023practices}{Yonadav Shavit
  and others, {``Practices for Governing Agentic AI Systems''}
  {[}2023{]} Research Paper, OpenAI}.}

A prime example is OpenAI's DeepResearch,\footnote{\url{https://openai.com/index/introducing-deep-research/}}
which can autonomously conduct comprehensive research, moving beyond
simple queries to planning multi-step investigations, analyzing data
from diverse sources, and synthesizing findings into detailed, cited
reports---tasks previously requiring the expertise and judgment of human
analysts. DeepResearch makes independent decisions about which sources
to trust, how to weigh conflicting information, and how to structure its
final report, thereby illustrating the creative decision-making that
characterizes agentic AI.\footnote{\citeproc{ref-acharya2025agentic}{Deepak
  Bhaskar Acharya, Karthigeyan Kuppan and B Divya, {``Agentic AI:
  Autonomous Intelligence for Complex Goals--a Comprehensive Survey''}
  {[}2025{]} IEEE Access}.} Yet, the overarching need for the research,
the direction of the research, and the utilization of the research
remain in the domain of the user. AI here is more than an
amanuensis\footnote{\citeproc{ref-ginsburg2019authors}{Jane C Ginsburg
  and Luke Ali Budiardjo, {``Authors and Machines''} (2019) 34 Berkeley
  Tech. LJ 343}.} but less than a collaborator---it makes decisions that
relate to the form of the outcome, but does not provide the motivation
for the research or shape its use.\footnote{The characterization of AI
  as \emph{amanuensis} is consistent with historical legal treatment of
  technological aids. Courts and copyright offices have traditionally
  viewed such tools---from cameras to word processors---as extensions of
  human agency rather than independent creators. \emph{See}, e.g., U.S.
  Copyright Office, \emph{Compendium of U.S. Copyright Office Practices}
  § 313.2 (3d ed.~2021) (``The Office will not register works produced
  by a machine or mere mechanical process that operates randomly or
  automatically without any creative input or intervention from a human
  author''). This framework positions AI systems as modern equivalents
  of scribal tools, executing tasks under human direction.}

Central to this distinction between generative AI and agentic AI is a
shift from a reactive, advisory role to proactive execution. Unlike
traditional generative AI, which responds to user prompts, agentic AI is
designed to tackle open-ended tasks extending beyond its initial
training data. It is distinguished by a capacity to emulate human-like
reasoning and communication, enabling it to plan strategies, adapt
dynamically to unforeseen conditions, and generate novel solutions
expressed in natural language---capabilities often seen as bridging into
human-level judgment calls.\footnote{The European Union's AI Act
  (Regulation (EU) 2024/1689) adopts a risk-based approach to regulating
  AI systems. Systems deemed `high-risk,' potentially including some
  applications of agentic AI due to their capacity for autonomous
  decision-making, are subject to stringent requirements regarding
  transparency, data governance, and human oversight.} This proactive
orchestration allows agentic AI to coordinate with other agents or
humans to achieve complex objectives. For instance, agentic AI could
autonomously negotiate pricing with suppliers in global supply chains,
dynamically reroute shipments to avoid geopolitical disruptions, and
recalibrate production schedules in response to fluctuating demand,
potentially reshaping operational workflows across industries.

Recent scholarship elaborates on this distinction. Shavit and
others\footnote{\citeproc{ref-shavit2023practices}{(N 5)}.} define
agentic AI as systems that ``pursue complex goals with limited direct
supervision,'' underscoring the leap from rigid, rule-bound intelligent
agents to AI that can self-initiate multi-step strategies. Similarly,
Caballar\footnote{\citeproc{ref-caballar2024aiagents}{(N 4)}.}
emphasizes that while traditional agents require constant human
oversight and operate under fixed protocols, agentic AI is characterized
by an ability to interpret context, execute complex plans, and adjust
strategies on the fly, marking a fundamental shift in how AI systems
interact with the world.

However, while this shift marks a novel form of digital
agency,\footnote{Agentic AI systems derive their behavior from RL
  processes that optimize for reward signals designed to align with
  human decision-making patterns and successful outcomes, rather than
  from intrinsic motivation. Consequently, their agency is
  functional---emulating human-like behavioral patterns---yet lacks the
  conscious intentionality that characterizes genuine human agency.} it
falls short of true autonomy. Franklin and Graesser\footnote{\citeproc{ref-franklin1996agent}{(N
  1)}.} characterize a truly autonomous agent as ``a system situated
within and a part of an environment that senses that environment and
acts on it, over time, in pursuit of its own agenda and so as to effect
what it senses in the future,'' a definition maintained by Bartosz
Brożek and Marek Jakubiec\footnote{\citeproc{ref-brozek2017legal}{{``On
  the Legal Responsibility of Autonomous Machines''} (2017) 25
  Artificial Intelligence and Law 293}.}. While agentic AI can act
autonomously in carrying out tasks, it still operates under the goals
and constraints set by its human users.\footnote{While Agentic AI's
  capacity to autonomously pursue long-term goals, make decisions,
  execute complex workflows, and proactively orchestrate multi-step
  processes, presents a novel agency, its degree of independence from
  direct human control in executing those actions is far more limited
  than complete (true) autonomy.} Therefore, it remains distinct from
(hypothetical) artificial general intelligence (AGI), which could
theoretically be endowed with independent will, a prerequisite often
considered for legal personhood, though this remains a contentious
issue.\footnote{\citeproc{ref-bryson2017and}{Joanna J Bryson, Mihailis E
  Diamantis and Thomas D Grant, {``Of, for, and by the People: The Legal
  Lacuna of Synthetic Persons''} (2017) 25 Artificial Intelligence and
  Law 273}.}

The (partial) autonomy of agentic AI is stochastic, dynamic, and
fluid---qualities that fundamentally distinguish it from both
traditional AI and (hypothetical) AGI. Unlike symbolic AI, which is
explicitly programmed with deterministic rules, modern agentic AI is
constructed using generative AI and trained implicitly via RL. This
model structure and training method introduces \emph{stochasticity} (the
AI's behavior is not predetermined): while some actions follow directly
from user instructions, others emerge from the AI's adaptive processes
(based on prior user inputs and feedback), and others from its
training.\footnote{The stochastic nature of agentic AI stems from its
  reliance on generative models and reinforcement learning, using
  mechanisms such as epsilon-greedy exploration, where the agent
  randomly selects a non-optimal action with probability \(\epsilon\) to
  explore alternative strategies. This introduces intrinsic variability,
  making the agent's optimization path non-deterministic. \emph{See},
  e.g., Richard S Sutton, Andrew G Barto, and others,
  \citeproc{ref-sutton1998reinforcement}{\emph{Reinforcement Learning:
  An Introduction}, vol 1 (MIT press Cambridge 1998)}.} Its autonomy is
also \emph{fluid}, influenced by the user's overarching objectives and
the evolution of the creative process, with some goals arising
autologously (i.e., in response to its own prior outputs). Furthermore,
it is \emph{dynamic}, as the user's inputs are shaped by the AI's prior
responses, and the AI's actions are influenced by the user's prior
requests.

Consequently, predicting how an AI agent will carry out a request---and
the extent to which it will follow the user's instructions rather than
asserting its autonomy---becomes difficult, if not impossible. For
example, an AI research agent might autonomously use different sources
than those specified by the user, even contradicting explicit guidance.
Such concerns do not arise with standard LLMs and traditional AI, which
lack autonomy, and are distinct from those posed by truly autonomous
systems, where the user is effectively irrelevant.

These factors introduce novel challenges for legal and policy
frameworks. The issue at hand is not merely how we conceptualize human
and machine contributions---significant progress has been made on such
questions---but rather the fundamental \emph{unmappability} of roles and
contributions within intertwined human-machine creative
processes.\footnote{Some scholars, such as Katie D Evans, Scott A
  Robbins and Joanna J Bryson, \citeproc{ref-evans2023we}{{``Do We
  Collaborate with What We Design?''} {[}2023{]} Topics in Cognitive
  Science}, argue against framing human--machine interactions as
  collaborative, asserting that true collaboration requires shared
  intentionality, moral agency, and the ability to co-determine
  objectives---attributes they deem absent in AI, which they view as
  heteronomous tools (i.e., governed externally rather than by
  self-determination). However, this critique presumes that machines
  lack the autonomy to participate in open-ended creative processes.
  Agentic AI subverts this premise. For example, when tasked with
  producing a climate report, the AI might autonomously refocus the
  analysis from mitigation costs to adaptation ethics based on its
  assessment of emerging scholarship. While the human sets the broad
  mandate, the AI dynamically determines the specific objectives and
  methodological trajectory---a form of \emph{procedural
  co-determination} that blurs the intentional hierarchy (namely, that
  humans have intentions while a mere tool does not) central to
  heteronomy critiques. This fluid renegotiation of sub-goals defies
  clean attribution, making `collaboration' less a metaphor than a
  functional descriptor of their creative entanglement.} Contributions
in a work may defy categorization as originating solely from human or
machine sources.\footnote{Authors such as Annemarie Bridy,
  \citeproc{ref-bridy2012coding}{{``Coding Creativity: Copyright and the
  Artificially Intelligent Author''} {[}2012{]} Stan. Tech. L. Rev. 5}
  have considered the case where ``digital works (i.e., software
  programs) will, relatively autonomously, produce other works that are
  indistinguishable from works of human authorship'' (p.~3). However,
  they have maintained the assumption that the contributions of human
  and machine can be separated. This holds, for example, when the output
  in question is developed using an AI whose autonomy is predictable
  (e.g., a text-to-image AI will always generate a different output---a
  distinct image---but will always undertake the same action---it will
  generate an image) or even negligible. In contrast, agentic AI brings
  to the fore scenarios where, in addition to an AI's outputs being
  potentially indistinguishable from human outputs, its `collaborations'
  with human users are not \emph{disentanglable}.} In such cases, we
argue, frameworks may need to treat human and machine contributions
equivalently---not because of their inherent moral or economic equality,
but due to the practical impossibility of determining origin.

We organize our paper as follows. We begin by establishing the
stochastic, fluid, and dynamic autonomy that characterizes modern
agentic AI systems, analyzing how their outputs---emerging from both
prior and immediate user inputs and AI outputs---recursively adapt
across multiple interactions. We then examine the implications for
authorship frameworks, grounding our analysis primarily in U.S.
Copyright law to maintain focus. Subsequent sections address
inventorship challenges in patent law and liability allocation in tort
frameworks. Throughout these domains, we trace how the fundamental
inability to disentangle human and AI contributions destabilizes current
legal and policy paradigms, identifying critical pressure points for
decision-makers. We conclude that conventional distinctions between
human and AI creations may require reconfiguration to address the unique
challenges posed by agentic AI systems.

\section{Stochastic, Dynamic, and Fluid Autonomy in Agentic
AI}\label{stochastic-dynamic-and-fluid-autonomy-in-agentic-ai}

Prevailing discourse on AI authorship, inventorship, and liability often
relies on a binary conceptualization of AI autonomy.\footnote{Daniel J
  Gervais, \citeproc{ref-gervais2019machine}{{``The Machine as Author''}
  (2019) 105 Iowa L. Rev. 2053} presents a related taxonomy where, at
  one extreme, stand machines like word processors that can be
  considered mere tools, and at the other, machines such as video games
  where the user merely selects between programmed options (p.~2069). He
  later rejects this classification for modern AI (`deep learning
  machines'), in part recognizing AI's capacity for autonomy, but from
  the perspective of the unpredictability (stochasticity) of the
  machine's outputs and its ability to develop high-level
  representations (e.g., Word2Vec) that capture correlations in the
  data. He does not discuss the dynamic adaptability and the contextual
  autonomy (i.e., autonomy that varies depending on user instructions
  and the specific task context) that characterize modern agentic AI.}
At one pole lies traditional generative AI, where users maintain almost
complete control over the AI's actions through iterative prompting and
output curation.\footnote{\citeproc{ref-samuelson2023generative}{Pamela
  Samuelson, {``Generative AI Meets Copyright''} (2023) 381 Science
  158}.} For example, text-to-image systems like DALL-E generate outputs
conditioned on human-provided prompts, with any creative variation
constrained by the input parameters. In each iteration, the AI generates
an image---the AI's output may be unpredictable, but its \emph{action}
is predictable. A human user might experiment with different prompts and
then curate the AI-generated images, selecting the most desirable ones.
At the other pole lies hypothetical AGI, capable of \emph{sovereign
autonomy}---independently conceiving and executing creative agendas
without any human oversight or direction.\footnote{\citeproc{ref-nick2014superintelligence}{Bostrom
  Nick, {``Superintelligence: Paths, Dangers, Strategies''}}.}

Agentic AI disrupts this dichotomy by introducing a partial autonomy
(intermediate between the extremal poles of negligible and complete
autonomy) that is stochastic, dynamic, and fluid. This autonomy is
partial because while the AI may exercise significant autonomy in
execution, it still operates under human-defined parameters: the
overarching objectives and constraints are set by a human user (e.g.,
``write a research paper, citing only peer-reviewed papers'' or
``optimize supply chain costs without introducing new vendors''). This
is important because were the autonomy either negligible or complete,
attributing contributions would be more clear-cut: with negligible
autonomy (i.e., if AI were merely a tool), all creative output would be
attributable to the human; with complete autonomy, the AI would be the
sole creator.

This autonomy is \emph{stochastic} because, unlike symbolic AI, which is
programmed with explicit, deterministic rules, modern agentic AI is
built upon generative AI. Its internal intermediate steps (the course of
its analysis) and its final outputs are both probabilistic. The precise
extent of the AI's autonomy is not fully predetermined or controllable
by the user and depends on the stochastic process it has learned to
emulate, relating prior inputs and outputs to subsequent
outputs.\footnote{\emph{See}, e.g., in the context of variational
  autoencoders, Diederik P Kingma, Max Welling, and others,
  \citeproc{ref-kingma2019introduction}{{``An Introduction to
  Variational Autoencoders''} (2019) 12 Foundations and
  Trends{\textregistered} in Machine Learning 307}; generative
  adversarial networks, Ian J Goodfellow and others,
  \citeproc{ref-goodfellow2014generative}{{``Generative Adversarial
  Nets''} (2014) 27 Advances in neural information processing systems};
  diffusion models Jonathan Ho, Ajay Jain and Pieter Abbeel,
  \citeproc{ref-ho2020denoising}{{``Denoising Diffusion Probabilistic
  Models''} (2020) 33 Advances in neural information processing systems
  6840}; and autoregressive models (like GPT) Tom Brown and others,
  \citeproc{ref-brown2020language}{{``Language Models Are Few-Shot
  Learners''} (2020) 33 Advances in neural information processing
  systems 1877}. In these architectures, the neurons learn transfer
  functions that combine low-level representations into increasingly
  abstract, high-level representations
  (\citeproc{ref-bengio2007scaling}{Yoshua Bengio, Yann LeCun, and
  others, {``Scaling Learning Algorithms Towards AI''} (2007) 34
  Large-scale kernel machines 1}). The weights and biases of these
  neurons, learned during training, define the parameters of the
  probability distribution that governs the stochastic generation
  process.} Thus, while the user sets the overall objective, the AI's
internal processes and decision-making pathways can lead to varying
autonomous actions. For example, two researchers using DeepResearch to
analyze ``climate policy efficacy'' might receive structurally distinct
reports: one emphasizing econometric modeling because the AI uncovered
relevant economic journal articles, and another prioritizing
sociopolitical feasibility because the system identified pertinent
policy journals. These reports may differentially align with the task
specified by the user, and indeed even with the user's goals.\footnote{The
  outputs of agentic AI can exhibit chaotic divergence due to compounded
  stochasticity across its recursive workflows. Unlike single-step
  generative systems (e.g., DALL-E's image variations from static
  prompts), agentic AI introduces randomness at three interdependent
  levels: (1) probabilistic action selection at each decision node, (2)
  path-dependent adaptation to prior workflow states, and (3)
  interpretive variance in processing user feedback. This creates the
  computational analog of the ``butterfly effect,'' where microscopic
  differences in initial conditions can lead to macroscopic outcome
  divergence. For instance, an agent analyzing climate policy might
  bifurcate into econometric or sociopolitical frameworks based on early
  source selection---a probabilistic choice during initial literature
  review that then recursively biases all subsequent analysis.}

It is \emph{fluid} because modern manifestations of agentic AI are
distinct from prior conceptualizations, as they are trained to learn
implicitly rather than being programmed explicitly. For example, when
considering machine creativity, Ginsburg and Budiardjo\footnote{\citeproc{ref-ginsburg2019authors}{(N
  8)}.} write, ``The computer scientist who succeeds at the task of
`reduc{[}ing{]} {[}creativity{]} to logic' does not generate new
`machine' creativity---she instead builds a set of instructions to
codify and simulate `substantive aspect{[}s{]} of human {[}creative{]}
genius,' and then commands a computer to faithfully follow those
instructions.'' Inherent in this conceptualization is the idea that the
AI was programmed to be creative rather than learning to be creative.
While the former was true for symbolic AI systems, modern agentic AI
learns from data and interactions with users. Its programmers do not
explicitly code instructions for the AI to follow; rather, the AI learns
to be creative through mechanisms like positive and negative
reinforcement. This RL training makes modern agentic AI's behavior
contextual (adaptive to user instructions), resulting in a fluid
autonomy where the level of human control is less clearly defined and
subject to change during operation---its planning, execution, and
outputs can vary significantly across different interactions and
tasks.\footnote{Consequently, agentic AI can exhibit emergent
  behavior--complex, unpredictable patterns that result from its
  training, inference, and model structure
  (\citeproc{ref-mitchell2009complexity}{Melanie Mitchell,
  \emph{Complexity: A Guided Tour} (Oxford university press 2009)}).
  While symbolic AI could exhibit some unexpected behaviors due to the
  complexity of its rules, the scale and nature of emergent behavior in
  agentic AI, driven by its learning mechanisms, are qualitatively
  different.}

It is \emph{dynamic} because the AI responds to current user inputs
within the context of prior user inputs, feedback, and its own previous
responses. This creates a feedback loop: the AI's autonomy in a given
interaction is shaped by its prior autonomy and the user's response to
it. Negative user feedback on excessive autonomy may lead the AI to
curtail it, while positive feedback may encourage greater initiative.
The user's guidance shapes the AI's autonomy, influencing the balance
between following specific directions and exercising independent
assessments.

Agentic AI's stochastic, fluid, and dynamic autonomy manifests along
three key, interconnected dimensions. \emph{Temporally}, human oversight
dominates during initial goal-setting, while the AI assumes increasing
control during execution phases. \emph{Functionally}, humans define
strategic objectives, while the AI operationalizes them through
context-sensitive decisions. \emph{Adaptively}, the system modifies its
creative approach based on user feedback, both implicit and explicit.
For instance, consider a user who sets a research AI agent (such as
DeepResearch) the strategic objective of assessing the ethical
implications of AI-driven diagnostics. \emph{Temporally}, the user
initiates the research by defining this broad topic, while the agent
independently manages the subsequent execution, from identifying
relevant publications to synthesizing findings into a structured report.
\emph{Functionally}, the agent autonomously determines the appropriate
analytical frameworks, potentially choosing to compare different ethical
guidelines across various countries---a level of detail not explicitly
specified by the user. \emph{Adaptively}, the agent refines its approach
based on user feedback; for example, if a user consistently prioritizes
peer-reviewed articles over preprints, the system will learn to favor
such sources in future research endeavors, even without direct
instruction, effectively internalizing the user's scholarly
preferences.\footnote{This adaptivity can arise through several
  complementary mechanisms. First, \emph{in-context learning} enables
  the system to draw upon prior interactions---such as user prompts and
  the model's own outputs---that remain within its context window, the
  portion of input data currently accessible during inference. Second,
  \emph{implicit preference learning}, often implemented through
  reinforcement learning techniques, allows the model to adjust its
  behavior based on patterns of user approval or correction over time.
  Third, \emph{explicit adaptation} may occur either through direct user
  instruction or via fine-tuning, where the underlying model parameters
  are updated based on aggregated user feedback. Fourth, during
  \emph{retrieval-augmented generation}, the system can dynamically
  prioritize external information sources---such as favoring
  peer-reviewed articles over preprints---in response to inferred or
  specified preferences. Together, these mechanisms contribute to a form
  of `relationship memory' that evolves across interactions.}

Crucially, this adaptivity creates recursive feedback loops---processes
where outputs in prior interactions become inputs in subsequent
interactions---between AI and human.\footnote{A recursive co-evolution
  of human and AI contributions finds conceptual analogs in several
  social science theories. Actor-network theory (ANT), which rejects
  hierarchical distinctions between human and non-human ``actants''
  (\citeproc{ref-latour2005reassembling}{Bruno Latour,
  \emph{Reassembling the Social: An Introduction to
  Actor-Network-Theory} (Oxford university press 2005)}), provides a
  particularly apt framework. ANT's symmetrical treatment of agency
  aligns with the paper's argument that human-AI creative entanglement
  defies traditional attribution. Similarly, Giddens' structuration
  theory (\citeproc{ref-giddens1984constitution}{Anthony Giddens,
  \emph{The Constitution of Society: Outline of the Theory of
  Structuration} (Univ of California Press 1984)})---where social
  structures and individual agency recursively shape one
  another---offers parallels to the fluid autonomy dynamics described
  here.} For instance, in the DeepResearch example, prioritizing
research sources, adjusting analytical methods, or replicating user
patterns may reflect learned responses to prior user feedback. A legal
scholar who previously emphasized comparative constitutional law in
their prompts may find the AI autonomously expanding its analysis to
include foreign jurisprudence in future projects---not because the user
explicitly requested it, but because the system has learned to amplify
and recombine the user's demonstrated preferences.

The interplay of these factors---stochastic variation, recursive
human-AI causal entanglements, and the adaptive nature of the AI's
autonomy---significantly complicates the determination of creative
contribution and control. For example, the AI may incorporate elements
from the human user's prior inputs and feedback into its outputs. In
such cases, its actions might be better characterized as a curation of
the human user's creativity, rather than independent
creation.\footnote{\citeproc{ref-bender2021dangers}{Emily M Bender and
  others, {``On the Dangers of Stochastic Parrots: Can Language Models
  Be Too Big?''} \emph{Proceedings of the 2021 ACM conference on
  fairness, accountability, and transparency} (2021)}.} This raises the
possibility that even an AI's ostensibly autonomous outputs could be
considered \emph{functionally} derivative.\footnote{This is based on the
  definition of derivative works as ``based upon one or more preexisting
  works'' through recasting, transformation, or adaptation (17 U.S.C. §
  101). However, this application enters a doctrinal gray zone. Unlike
  traditional derivative works where human authors consciously intend to
  create a new work based on a pre-existing one, here the AI's
  adaptation occurs through probabilistic inference from historical
  interactions. This process lacks the \emph{mens rea}, or mental state,
  typically associated with copyright authorship. Thus, a core issue is
  whether an AI process can even create a derivative work in the legal
  sense, in the absence of human authorial intent. Moreover, under 17
  U.S.C. § 106(2), AI cannot be recognized as an author. Therefore, even
  its human user's intent to create a derivative work, if present, might
  not be sufficient if the AI is deemed the primary source of the new
  expression.} Given this blending of human and machine contributions,
is the resulting work a product of human intent, machine autonomy, or an
inseparable fusion of both? The AI's outputs, shaped by unpredictable
probabilistic processes and prior user instructions, resist clear
attribution.

\section{Authorship}\label{authorship}

Scholars have long questioned whether traditional copyright
frameworks---built around the notion of the human creator---can fully
capture works generated by algorithmic processes.\footnote{\citeproc{ref-samuelson1985allocating}{Pamela
  Samuelson, {``Allocating Ownership Rights in Computer-Generated
  Works''} (1985) 47 U. pitt. L. rev. 1185};
  \citeproc{ref-acosta2012artificial}{Raquel Acosta, {``Artificial
  Intelligence and Authorship Rights''} (2012) 17 Harvard Journal of Law
  and Technology}; \citeproc{ref-jaszi2017toward}{Peter Jaszi, {``Toward
  a Theory of Copyright: The Metamorphoses of {`Authorship',}''}
  \emph{Intellectual property law and history} (Routledge 2017)};
  \citeproc{ref-abbott2020reasonable}{Ryan Abbott, \emph{The Reasonable
  Robot: Artificial Intelligence and the Law} (Cambridge University
  Press 2020)}.} At the heart of this debate lies a central question:
When AI generates the intellectual content, \emph{who is the author}?
And, flowing from that, \emph{who owns the copyright}? Could it be the
artist or writer who commissioned the work, the AI service provider who
built the system, the AI itself, or perhaps no one at all?

This human-centric paradigm faces mounting theoretical challenges.
Bridy\footnote{\citeproc{ref-bridy2012coding}{(N 20)}.}, for example,
challenges the entrenched assumption of uniquely human authorship by
arguing that creativity itself is inherently algorithmic. She
illustrates that even what we typically consider ``human'' creativity
operates through rules and structured processes, suggesting that works
produced autonomously by computers are less alien to our creative
paradigms than conventional law presumes. Her analysis underscores that,
if the law is to remain relevant in an era increasingly defined by AI,
it must evolve beyond its narrow human-centric lens to accommodate the
new realities of machine-generated creative output.

However, current legal frameworks remain fundamentally anthropocentric,
hinging on whether a human has exercised meaningful control over
AI-generated outputs---a benchmark that increasing AI autonomy
complicates.\footnote{\citeproc{ref-zeilinger2021tactical}{Martin
  Zeilinger, \emph{Tactical Entanglements: AI Art, Creative Agency, and
  the Limits of Intellectual Property} (meson press 2021)}.} This is
exemplified by the U.S. Copyright Office's 2023 policy, which affirms
that AI-generated works lacking substantial human authorship cannot be
copyrighted, thereby creating significant ambiguity regarding protection
and ownership, particularly in cases of intertwined human and machine
contributions.\footnote{\citeproc{ref-lemley2023generative}{Mark A
  Lemley, {``How Generative AI Turns Copyright Law on Its Head''}
  {[}2023{]} Available at SSRN 4517702}.}\footnote{U.S. Copyright
  Office, \emph{Copyright Registration Guidance: Works Containing
  Material Generated by Artificial Intelligence}, 88 Fed. Reg. 16190
  (Mar.~16, 2023).} For instance, the Office refused registration for
the AI-generated comic \emph{Zarya of the Dawn}, holding that the human
user's prompts (e.g., ``adjust lighting,'' ``make the tiger look more
menacing'') were insufficiently creative to constitute authorship,
effectively treating the AI as a ``tool'' rather than a
collaborator.\footnote{U.S. Copyright Office, Letter to Van Lindberg,
  Esq., Re: Zarya of the Dawn (Registration \# VAu001480196) (Feb.~21,
  2023).} In stark contrast, Chinese courts have taken a more expansive
view.\footnote{Chinese courts have offered contrasting perspectives on
  AI authorship. In \emph{Beijing Film Law Firm v. Beijing Baidu Netcom
  Science \& Technology Co., Ltd.}, {[}2018{]} Jing 0491 Min Chu No.~239
  (Beijing Internet Ct. Apr.~25, 2019), the Beijing Internet Court held
  that only works created by natural persons qualify for copyright
  protection under Chinese law, thus denying protection to output
  generated by computer software, even if original. However, in
  \emph{Shenzhen Tencent Computer System Co., Ltd.~v. Shanghai Yingxun
  Technology Co., Ltd.}, {[}2019{]} Yue 0305 Min Chu 14010 (Shenzhen
  Nanshan Dist. People's Ct. Dec.~24, 2019), the Nanshan District Court
  of Shenzhen took a different approach. It granted copyright protection
  to an article generated by the ``Dreamwriter'' software, emphasizing
  the human involvement in selecting and arranging the data and
  parameters that shaped the AI's output. This decision recognized that,
  while the AI generated the text, the human contribution to the overall
  process was sufficient to meet the requirements for a ``written work''
  under Chinese copyright law. For a detailed discussion, see Yong Wan
  and Hongxuyang Lu, \citeproc{ref-wan2021copyright}{{``Copyright
  Protection for AI-Generated Outputs: The Experience from China''}
  (2021) 42 Computer Law \& Security Review 105581}.} As exemplified by
the \emph{Shenzhen Tencent} decision, the court granted copyright
protection to an AI-generated news article, emphasizing the human
involvement in curating training data, selecting input variables, and
setting system parameters---activities that, while arguably less direct
than the prompting in \emph{Zarya}, were deemed sufficient to establish
authorship under Chinese law. This divergence highlights a fundamental
tension: Is direct, expressive input (like detailed prompting) the
\emph{sine qua non} of authorship, or can more indirect, preparatory
contributions suffice?

Critically, these debates---concerning authorless versus authored
works---\footnote{\citeproc{ref-lemley2023generative}{Lemley (n 35)}.}and
proposed solutions---such as hybrid attribution models,\footnote{Co-listing
  human and AI contributors, \citeproc{ref-abbott2020artificial}{R
  Abbott, {``Artificial Intelligence, Big Data and Intellectual
  Property: Protecting Computer Generated Works in the United
  Kingdom,{[}w:{]} Research Handbook on Intellectual Property and
  Digital Technologies, Red''} {[}2020{]} T. Aplin,
  Cheltenham--Northampton}.} two-tiered protection systems,\footnote{\citeproc{ref-sun2021redesigning}{Haochen
  Sun, {``Redesigning Copyright Protection in the Era of Artificial
  Intelligence''} (2021) 107 Iowa L. Rev. 1213}.} or Gervais's theory of
`originality causation'---\footnote{\citeproc{ref-gervais2019machine}{Gervais
  (n 21)}.}assume the ability to \emph{parse} the contributions of human
and AI. For instance, \emph{if} human and AI contributions \emph{could}
be clearly delineated, a work could potentially be recognized as a
collaborative creation.\footnote{Whether this is advisable is another
  question, with arguments falling on both sides (see, e.g.,
  \citeproc{ref-grimmelmann2015there}{James Grimmelmann, {``There's No
  Such Thing as a Computer-Authored Work-and It's a Good Thing, Too''}
  (2015) 39 Colum. JL \& Arts 403} (arguing against AI authorship);
  \citeproc{ref-sun2021redesigning}{Sun (n 41)} (proposing sui generis
  rights for AI-generated works with human inputs)).} This might involve
crediting the human author for creative direction and either
acknowledging the AI's role in a new category (e.g., ``AI-assisted
creation'') or attributing the AI-generated portions to the human by
extension. Alternatively, dynamic royalty schemes could be adopted:
instead of asking ``who is the author?'', the focus could shift to ``how
much is each an author? Who should benefit, and how much?''. A song
generated by AI, for example, could trigger a royalty allocation among
the human who commissioned it, the AI's developer, and potentially a
fund for creators whose works trained the AI.\footnote{The EU Data Act
  (Regulation (EU) 2023/2854) addresses data access and sharing, with
  provisions on fair compensation for data generation, but does not
  directly address AI training or output royalties. The EU AI Act
  (Regulation (EU) 2024/1689) regulates AI systems, including data
  governance, but similarly lacks specific provisions on output
  royalties, though its broader implications for copyright are subject
  to analysis (see, e.g., \citeproc{ref-quintais2025generative}{João
  Pedro Quintais, {``Generative AI, Copyright and the AI Act''} (2025)
  56 Computer Law \& Security Review 106107}).} These royalties could be
adjusted based on relative contributions: a human who heavily edited the
AI's output would receive a larger share, while a largely AI-generated
work might favor the developer. Another option involves considering
\emph{sui generis} rights---limited protections weaker than full human
authorship but stronger than the public domain.\footnote{Unlike
  proportional royalties, which operate within existing copyright
  frameworks to distribute revenue, \emph{sui generis} rights create a
  new framework with its own rules for protection, duration, and scope.
  Existing \emph{sui generis} regimes like the EU Database Directive
  (96/9/EC), protecting non-creative investments (e.g., data
  compilation) for 15 years
  (\citeproc{ref-reichman1997intellectual}{Jerome H Reichman and Pamela
  Samuelson, {``Intellectual Property Rights in Data''} (1997) 50 Vand.
  L. Rev. 49}), may offer a precedent for AI-generated works. This
  approach avoids the need to determine ``authorship'' in the
  traditional sense, focusing instead on the outcome (the AI-generated
  work) and granting limited rights based on technical criteria (e.g.,
  evidence of AI synthesis) rather than human creative input. A key
  advantage is sidestepping the attribution problem, but a risk is
  potentially incentivizing a flood of AI-generated content, potentially
  impacting the value of human-created works.}

Without the ability to reliably parse contributions, however, these
questions, debates, and proposed solutions become largely moot. While
attributing distinct human and AI inputs may remain feasible in some
straightforward settings---thus permitting conventional legal standards
to apply---the real challenge arises with the continuum of recursive
agentic AI-human interactions where contributions become increasingly
entangled. Specifically, any framework premised on distinguishing the
origin of creative elements faces two intractable problems: (1) ensuring
fair and consistent treatment across cases where contributions are
separable versus those where they are inseparable, and (2) establishing
reliable criteria for determining whether contributions can even be
parsed in the first place.

Consider two classes of works resulting from human-AI interaction. Works
in the first (separable) class allow specific creative elements to be
reasonably attributed \emph{post hoc} to either the human or the AI. For
example, the human might have written distinct sections while the AI
generated others, or clear logs might delineate contributions. In the
second (inseparable) class, the interaction, likely involving recursive
feedback loops, results in an inextricably blended work---a fusion where
the origins of specific ideas, phrasings, or creative choices are
fundamentally entangled and untraceable.

This division yields a dilemma for any single, attribution-based legal
standard. On the one hand, frameworks requiring the separation of human
and AI contributions (e.g., granting full copyright only to
human-generated portions) immediately fail when applied to the
inseparable class, as the necessary distinctions cannot be made. On the
other hand, a framework suitable for inseparable works must operate
\emph{without} assessing the extent of specific contributions. Such a
framework, if applied to the separable class, could not account for
variations in human versus AI input, treating works with potentially
vastly different contribution levels identically. Therefore, it is
impossible to create a single, attribution-based standard that both
functions for inseparable works and appropriately differentiates between
separable works based on contribution levels.

Suppose instead we developed two distinct standards, one tailored for
separable works and another for inseparable ones. The immediate
challenge shifts to reliably determining whether a specific work belongs
to the `separable' or `inseparable' class. Along the continuum of
human-AI interaction, making this determination---deciding whether
contributions are truly separable or inextricably fused---is likely to
be often subjective and prone to inconsistency. How should works be
treated where \emph{some} but not all elements might be attributable?
Does the presence of \emph{any} inseparable element necessitate
classifying the entire work as inseparable? If so, a vast majority of
works involving recursive agentic AI interaction might fall into the
inseparable category, rendering the `separable' standard largely
irrelevant in practice. Moreover, how could we ensure that these two
distinct standards yield equivalent results? Without such equivalence,
works reflecting similar human effort could receive different legal
treatment based merely on the traceability of the creative process and
not its substance.

These challenges are greatly amplified by \emph{recursive
adaptation}\footnote{When AI systems adapt their creative processes
  based on human user feedback, and human users adapt their creative
  processes based on AI feedback. This produces a \emph{creative
  ouroboros}---a self-referential loop where human and machine
  contributions mutually reconstitute each other across iterative
  cycles.} in human-agentic AI interactions. Consider an AI graphic
designer agent that evolves its artistic style to align with a human
client's historical preferences and inputs, absorbing and fine-tuning
its outputs based on the user's inputs and interactions. Suppose, also,
that the human client evolves her style to match the AI's outputs,
learning from the AI.\footnote{For instance, linguistic alignment, also
  known as convergence, is a well-established concept in
  psycholinguistics, where conversational partners tend to mimic each
  other's language use, including word choices, phrasings, and syntactic
  structures (\citeproc{ref-pickering2004toward}{Martin J Pickering and
  Simon Garrod, {``Toward a Mechanistic Psychology of Dialogue''} (2004)
  27 Behavioral and brain sciences 169}). In human-AI interactions, this
  concept implies users may adapt their language over time when
  interacting with artificial agents (e.g., see
  \citeproc{ref-vinchon2023artificial}{Florent Vinchon and others,
  {``Artificial Intelligence \& Creativity: A Manifesto for
  Collaboration''} (2023) 57 The Journal of Creative Behavior 472})}
This creates a \emph{causal entanglement} where neither the human user
nor the AI fully determines the creative trajectory; the AI system
itself becomes an active participant in the evolution of the human
designer's style, effectively curating prior human-AI interactions. In
such cases, how should rights be apportioned?

One might argue that an AI is merely a tool, incapable of autonomously
undertaking either derivative or transformative work. On this view, all
of the AI's outputs \emph{could potentially} qualify as authored works
\emph{by the user}, since everything produced by a tool (like a word
processor) is typically considered a reflection of its user's input.
Applied to agentic AI, this position would imply that outputs generated
by an agentic AI that adapted to its human users' inputs, guidance, and
previously authored works, may likewise be considered the user's
authored works.

However, now suppose the AI is capable of autonomous output. Further
suppose, for instance, that this agent generates output meeting
\emph{Feist}'s ``modicum of creativity'' standard\footnote{\emph{Feist
  Publications, Inc., v. Rural Telephone Service Co.}, 499 U.S. 340, 345
  (1991).} by internalizing and recombining its human user's prior
copyrighted works. Under current U.S. law, an AI cannot \emph{itself}
create derivative works, as only humans hold that capacity under 17
U.S.C. § 106(2). However, the human user's iterative feedback and
curation---even if insufficient on their own to meet the \emph{Feist}
standard for originality---might arguably establish a copyright claim to
the AI's output \emph{as} a derivative work based on the user's
underlying contributions.\footnote{A doctrinal frontier with no clear
  precedent.} This is because the AI's output could be seen as
\emph{functionally} derivative\footnote{The term `functionally'
  acknowledges that AI cannot legally create derivative works under 17
  U.S.C. § 106(2) as it is not recognized as an author. However, the
  AI's outputs may practically serve as derivatives of human creative
  inputs.} of the user's prior copyrighted works, which guided the AI's
adaptation.

This creates a paradox: if the AI's output is \emph{functionally}
derivative of the user's prior inputs, the human user may claim
authorship even if the AI operated autonomously and the user's specific
contributions during the interaction did not meet traditional authorship
requirements. That is, even if the human user did not meet the
requirements of the U.S. Copyright Office's 2023 guidance,\footnote{AI
  outputs ``determined primarily by the AI'' lack protection, but human
  authors may claim rights if they ``exercise creative control over the
  AI's output \emph{and} contribute original expression'' through
  iterative refinements.} and while the AI itself lacks authorship
rights, its output might still be subject to a claim of human authorship
asserted by the user based on derivative rights.\footnote{These issues
  are distinct from debates surrounding copyright and traditional
  generative AI, which primarily focus on whether the AI's output is
  substantially similar to its training data and whether the use of that
  data constitutes reproduction--essentially, whether the output is a
  functional derivative of the training data
  (\citeproc{ref-huang2025does}{Weijie Huang and Xi Chen, {``Does
  Generative AI Copy? Rethinking the Right to Copy Under Copyright
  Law''} (2025) 56 Computer Law \& Security Review 106100}). Agentic AI,
  in contrast, raises the question of whether its output is a functional
  derivative of its user's inputs.} This tension exists because while AI
legally cannot create derivative works under 17 U.S.C. § 106(2), the
human user might leverage the functional relationship between their
prior works and the AI's output to establish their claim to the new
work.

Moreover, even if the AI's use of its human user's inputs is
\emph{functionally} transformative,\footnote{AI cannot legally create
  transformative works under 17 U.S.C. § 106(2).} akin to a collage
artist transforming source material,\footnote{\emph{Cariou v. Prince},
  714 F.3d 694 (2d Cir. 2013). While \emph{Cariou} dealt with human
  appropriation of existing photographs, the underlying principle--that
  significant transformation of pre-existing material can create new
  copyrightable expression--is relevant to the AI context.} its outputs
may remain authorless yet be eligible for derivative authorship or
copyright protection by the human user. This is because, if the AI's
processes reflect the user's prior inputs and guidance, the user may be
positioned to claim rights over the resulting outputs, even if the
``creative spark''\footnote{``Creative spark'' denotes the originating
  creative idea or expressive choice that imbues a work with
  originality.} originated not from the human but from the AI's
autonomous generation. For example, the key motif in an output from an
AI-based graphic designer system might have emerged entirely from the AI
itself. Yet that resulting work could still be characterized as
authorless (since AI cannot legally be its own author) and
simultaneously subject to an authorship claim by the human user,
provided the motif emerged through the AI's assimilation of the user's
style and prior works.

Thus, the adaptive capabilities of agentic AI fundamentally challenge
current AI authorship doctrine. Furthermore, such recursive adaptation
destabilizes at least three other foundational doctrines.

First, the work-made-for-hire doctrine, codified in 17 U.S.C. § 201(b),
vests authorship in employers for works created by employees ``within
the scope of employment.''\footnote{17 U.S.C. § 201(b); Restatement
  (Third) of Agency § 7.07 (Am. L. Inst. 2006) (Employee Acting Within
  Scope of Employment). \emph{See also Cmty. for Creative Non-Violence
  v. Reid}, 490 U.S. 730 (1989) (establishing factors to determine
  employment status for work-made-for-hire).} \footnote{A significant
  difference exists between U.S. and European copyright law regarding
  works created within an employment context. The U.S.
  `work-made-for-hire' doctrine (17 U.S.C. § 201(b)) automatically vests
  copyright ownership in the employer. In contrast, many continental
  European jurisdictions, rooted in the concept of \emph{droit
  d'auteur}, initially vest copyright in the employee-creator (the
  natural person), with provisions for subsequent contractual transfer
  or licensing to the employer (\emph{See, e.g.}, German Copyright Act,
  Section 43; French Intellectual Property Code, Article L113-9).
  Notably, the UK's statutory approach differs from this continental
  model; under the Copyright, Designs and Patents Act 1988, Section
  11(2), the \emph{employer} is generally the first owner of copyright
  in works created by employees in the course of employment, subject to
  agreement. These differences in national legislations (\emph{see},
  Dénes Legeza, \citeproc{ref-legeza2015employer}{{``Employer as
  Copyright Owner from a European Perspective,''} \emph{SERCI annual
  congress 2015} (2015)} for a more detailed discussion of these
  variations within Europe), are likely to create further complexities
  in determining copyright ownership when employees in different
  countries employ distinct AI systems, potentially leading to
  conflicting claims of ownership.} This doctrine, however, presupposes
a human creator. If an agentic AI operates with a high degree of
autonomy (e.g., generating marketing copy without direct human
oversight), courts may reject work-made-for-hire claims because the AI
is neither an employee nor a legally recognized ``author.'' This creates
a gap: outputs generated by AI under broad corporate directives (e.g.,
``create a branding campaign'') may lack clear ownership, as no human
employee directly ``created'' the work.\footnote{ \emph{See, e.g.,} Mark
  A Lemley and Bryan Casey, \citeproc{ref-lemley2020fair}{{``Fair
  Learning''} (2020) 99 Tex. L. Rev. 743} (acknowledging the challenges
  AI-generated works pose to traditional copyright doctrines).} Yet, as
established earlier, agentic AI possesses only partial autonomy.
Therefore, if a human employee provides sufficient creative direction or
control over the AI's process, and the work is created within the scope
of their employment, the work-made-for-hire doctrine could potentially
still apply. The challenge lies in determining when human involvement
meets the threshold for ``sufficient creative direction,'' given the
AI's autonomous contributions. Courts assessing creative control often
examine who exercised ``superintendence'' over the work's creation, a
standard difficult to apply when a non-human agent contributes
significantly.\footnote{\emph{See Aalmuhammed v. Lee}, 202 F.3d 1227,
  1234--35 (9th Cir. 2000).}

Second, joint authorship standards, requiring intent to merge
contributions into a unitary whole\footnote{E.g., \emph{Childress v.
  Taylor}, 945 F.2d 500, 505--06 (2d Cir. 1991).} \footnote{Joint
  authorship is recognized across European copyright law, generally
  requiring a collaborative effort and a shared intention to create a
  unified work. National laws implementing Directive 2001/29/EC (the
  InfoSoc Directive) typically address joint authorship, although
  specific criteria and the rights of joint authors may vary. \emph{See,
  e.g.}, German Copyright Act, Section 8 (Joint Authors); French
  Intellectual Property Code, Article L113-2 (Work of Collaboration); UK
  Copyright, Designs and Patents Act 1988, Section 10 (pre-Brexit, but
  illustrative).}, are challenged when one potential ``author'' (the AI)
lacks legal personhood and the requisite intent. Agentic AI cannot form
the intent to collaborate when operating autonomously, particularly when
human user inputs are limited to high-level prompts (e.g., ``design a
logo in a retro style''). In such cases, courts may deem the AI
ineligible for joint authorship, even if its output reflects creative
synthesis. Furthermore, even in scenarios with involved human inputs and
feedback---and even where a human user explicitly asks the AI to
collaborate, potentially providing the human side of the intent---the
current legal framework does not recognize the AI as an entity capable
of forming or executing such intent. The fundamental problem of the AI's
lack of legal personhood persists, rendering it unable to meet joint
authorship requirements.

Third, in jurisdictions recognizing them, moral rights---such as the
right to attribution and the right to integrity of the work---are
inherently tied to the \emph{human} author's personal connection to
their creation.\footnote{Moral rights are a cornerstone of copyright law
  in many European jurisdictions, often stemming from the Berne
  Convention for the Protection of Literary and Artistic Works (Article
  6bis). These rights, typically including the right of attribution (to
  be identified as the author) and the right of integrity (to object to
  distortions of the work), are generally considered inalienable and
  remain with the author even after economic rights have been
  transferred. The specific scope and enforcement of moral rights vary
  across EU member states, but they generally provide significantly
  stronger protection for the author's personal connection to their work
  than in the U.S. \emph{See, e.g.}, German Copyright Act, Sections
  12-14 (Moral Rights) and French Intellectual Property Code, Article
  L121-1 (Moral Rights).} Agentic AI, lacking legal personhood, cannot
hold moral rights. However, the recursive interplay between human user
and AI agent complicates the attribution and protection of these rights
for the human user. When an AI significantly contributes to a work,
evolving its style and output based on the user's prior inputs and
feedback, the resulting creation becomes a blend of human and machine
agency. In these cases, attributing the work solely to the human user
becomes unclear, especially when the AI's autonomous contributions are
substantial. Furthermore, if the AI, through autonomous adaptation,
modifies the work in ways diverging from the human user's original
intent or artistic vision, the user's right to the \emph{integrity} of
the work may be challenged.\footnote{For instance, an AI literary agent
  might autonomously revise a manuscript to emphasize themes of
  algorithmic bias---a perspective the human author never explicitly
  endorsed but which emerged from the AI's analysis of their prior works
  on technology ethics. While the AI's alterations could enhance the
  work's social relevance, they simultaneously undermine the author's
  right to control the expression of their personal worldview.} Unlike
traditional scenarios where moral rights protect against derogatory
treatment by other humans, here the AI---employed by the human
user---autonomously alters the work, reflecting a novel conflict between
user control and AI agency.

\section{Inventorship}\label{inventorship}

The issues challenging authorship frameworks also arise in the context
of inventorship. A case in point is \emph{DABUS}, which involved an AI
system that generated novel inventions. Patent applications naming
\emph{DABUS} as the \emph{inventor}---directly challenging the
requirement of a \emph{human} conceiver---triggered legal battles
worldwide. Thus far, patent offices and courts in major jurisdictions
(U.S., U.K., EU) have rejected AI inventorship, insisting that inventors
must be natural persons.\footnote{In the U.S., the Patent and Trademark
  Office (USPTO) rejected the application in a decision dated April 22,
  2020 (Application No.~16/524,350), insisting that inventors must be
  natural persons. The U.S. Court of Appeals for the Federal Circuit in
  \emph{Thaler v. Vidal}, 43 F.4th 1207 (Fed. Cir. 2022), then affirmed
  that under current statutes, only humans can be inventors. The
  European Patent Office and UK Intellectual Property Office reached
  similar conclusions, also rejecting AI inventorship. However, there
  are notable outliers. South Africa granted a patent with DABUS as
  inventor, although this is seen as procedural rather than a legal
  endorsement due to their system's limited substantive review. In
  Australia, the Federal Court in \emph{Commissioner of Patents v.
  Thaler} {[}2021{]} FCA 879 initially ruled that AI could be an
  inventor, but this was unequivocally overturned on appeal by the Full
  Federal Court of Australia in \emph{Thaler v. Commissioner of Patents}
  {[}2022{]} FCAFC 62.}

The legal questions in the \emph{DABUS} case were relatively clear-cut
because no human participated in the conceptualization or design of the
inventions. How might the outcome differ if a human had played some
role, however minor, in the ideation or development process? One could
imagine a continuum from no human participation to solely human
participation, with AI systems potentially being fine-tuned or
development processes adjusted to facilitate human-AI partnerships
anywhere along that spectrum. At what point along this continuum would
we be willing to grant inventorship?\footnote{For example, contrast the
  varying decisions of the Chinese courts as discussed earlier, albeit
  in authorship, with the \emph{DABUS} case.} And critically, would
contributions even be separable at that juncture, making a standard
based on contribution levels practicable?\footnote{I.e., would we be
  able to measure contributions with sufficient accuracy at that point
  for such a standard to be practicable?}

Under U.S. patent law, inventorship requires both \emph{conception}
(``the complete performance of the mental part of the inventive act'')
and \emph{reduction to practice} (embodying the invention in a tangible
form).\footnote{ \emph{Burroughs Wellcome Co.~v. Barr Labs., Inc.}, 40
  F.3d 1223, 1227--28 (Fed. Cir. 1994).} Courts have long held that only
humans can conceive inventions, meaning only natural persons can be
legally recognized as inventors.\footnote{The European Patent Convention
  (EPC) also requires that an inventor be a natural person. Rule 19(1)
  EPC states that the request for grant of a European patent shall
  contain `the designation of the inventor.' The case law of the Boards
  of Appeal of the European Patent Office has consistently held that
  this designation must refer to a human being.} Agentic AI, however,
may autonomously `conceive'---or perhaps more accurately,
\emph{functionally conceive}---by generating novel solutions that
otherwise meet patentability criteria (e.g., non-obviousness,
utility).\footnote{However, the use of agentic AI raises further
  questions about \emph{demonstrating} that the obviousness standard is
  met. If an AI arrives at a solution that would be non-obvious to a
  human expert (a Person Having Ordinary Skill In The Art, or PHOSITA),
  but the AI's reasoning process is opaque (i.e., there is little
  evidence of the process that might support a claim of
  non-obviousness), how can one \emph{prove} that the solution meets the
  legal requirement?} \footnote{A further question concerns the meaning
  of obviousness in the context of autonomous AI. Given an innovation,
  if an AI could generate it when provided solely with prior information
  and overarching guidance, does this imply the innovation is obvious?
  If so, this standard arguably should also apply to human-generated
  innovations. Specifically, an innovation might be deemed obvious if an
  AI could reasonably generate it without specific human guidance, even
  if it was actually created by a human and appears non-obvious to human
  experts.}

For example, an AI drug discovery system might hypothesize and simulate
new molecular structures addressing a target disease mechanism---a
process traditionally constituting legal ``conception.'' Identifying the
extent to which an AI-generated invention draws upon its human user's
inputs and feedback would be critical to maintaining the human-only
conception requirement. As seen in the \emph{DABUS} case, if the AI
performed the core conception, the invention might lack a legally valid
conceiver, thereby failing a fundamental requirement for patentability
under current law. However, in scenarios where \emph{both} the human and
the AI `align' through recursive adaptation, the AI's adjustments based
on human inputs and feedback make it unclear whether the conception
originated with the human or the AI, thus obscuring who performed the
crucial ``mental part of the inventive act.''

The challenge extends to the second prong of inventorship:
\emph{reduction to practice}. This requires either physically embodying
the invention and demonstrating its utility (\emph{actual} reduction to
practice) or filing a patent application with a description sufficient
to enable a PHOSITA to make and use the invention (\emph{constructive}
reduction to practice) under 35 U.S.C. § 112(a). Agentic AI complicates
both pathways.

For \emph{actual} reduction to practice, AI systems integrated with
robotics or simulation tools can likely autonomously perform the
necessary physical steps or virtual testing. An AI might design,
synthesize, and test a novel compound without direct human intervention
in each step. However, if the AI executes these tasks based on a blend
of its own learned strategies (derived from recursive interactions),
direct human inputs, and autonomous decision-making, attributing the
successful reduction to practice becomes legally tenuous. Whose actions
ultimately demonstrated the invention worked for its intended purpose
when the process involves this blend of human guidance, recursive
adaptation, and autonomous AI execution?

The hurdles are perhaps even higher for \emph{constructive} reduction to
practice. While agentic AI can generate detailed technical descriptions
suitable for a patent draft, satisfying the enablement and written
description requirements of § 112(a) is fraught with difficulty.
Enablement demands that the disclosure teach a PHOSITA how to make and
use the invention without undue experimentation. If the AI's inventive
process relies on logic opaque to humans,\footnote{For instance,
  consider the evaluation functions in advanced chess engines (e.g.,
  Stockfish). These functions assign precise numerical scores to
  millions of board positions based on complex mathematical features,
  guiding vast computational searches. While this process is logically
  complete and demonstrably effective, its internal
  rationale---optimizing a complex mathematical function---differs
  fundamentally from human expert reasoning, which typically relies on
  strategic principles, pattern recognition, and established concepts
  (like named openings or positional advantages). Consequently, even if
  the AI's output (e.g., a novel chemical structure) is provided
  alongside the AI's code, the underlying inventive logic might remain
  opaque. A PHOSITA might not be able to understand or replicate the
  \emph{reasoning} leading to the invention using their field's
  conventional knowledge and techniques without undue experimentation,
  thus potentially failing the enablement requirement.} its generated
description might detail the outcome but fail to adequately explain the
underlying principles or non-obvious steps required for replication by a
human expert, potentially rendering the disclosure non-enabling. In
addition, the human user may be crucial in examining the AI's outputs to
ensure that the invention is sufficiently detailed for another human.
Could such iterative feedback constitute adequate guidance to claim
human inventorship?

Finally, the written description requirement necessitates showing the
\emph{human} inventor possessed the claimed invention \emph{at the time
of filing}. When an AI conceives the core idea and drafts the
description, demonstrating genuine human possession---beyond merely
receiving, understanding, and transmitting the AI's output---becomes
problematic. Did the human truly possess the invention in the legally
required sense if the complete mental conception originated
significantly with the AI, even if the human reviewed and filed the
AI-generated description? This challenges the fundamental link between
the human mind and the claimed subject matter required by the written
description doctrine.

Moreover, similar to the challenges identified in authorship, the
doctrine of joint inventorship faces distinct and novel difficulties
when confronted with agentic AI. Under current U.S. patent law, joint
inventors must each contribute significantly to the invention's
\emph{conception}---``the complete performance of the mental part of the
inventive act''---and typically engage in some form of collaborative
activity.\footnote{See \emph{Burroughs Wellcome Co.~v. Barr Labs.,
  Inc.}, 40 F.3d 1223, 1227--28 (Fed. Cir. 1994) (defining conception);
  \emph{Ethicon, Inc.~v. U.S. Surgical Corp.}, 135 F.3d 1456, 1460 (Fed.
  Cir. 1998) (requiring each joint inventor contribute to conception).
  \emph{See} also, \emph{Kimberly-Clark Corp.~v. Procter \& Gamble
  Distrib. Co.}, 973 F.2d 911, 917 (Fed. Cir. 1992) (indicating joint
  inventors usually collaborate or show connection, though contributions
  need not be equal nor efforts simultaneous). While the standard for
  collaborative intent in U.S. patent law may differ from the copyright
  standard articulated in \emph{Childress v. Taylor}, 945 F.2d 500 (2d
  Cir. 1991), the requirement for some joint effort remains. European
  frameworks, such as the European Patent Convention (EPC), generally
  concur, requiring contributions to the inventive concept from all
  collaborators.} Agentic AI disrupts this framework by introducing a
non-human entity capable of independently generating inventive concepts,
yet incapable of forming the requisite intent or holding legal status as
an inventor.

Consider an example from drug discovery: An AI system, guided by human
researchers, autonomously identifies a novel molecular structure
constituting the core inventive concept. The AI's contribution meets
technical criteria (novelty, utility), but it cannot be named an
inventor. Can the human researchers be named? If a \emph{single}
researcher merely provided high-level objectives, their contribution
might fail the conception standard. If \emph{multiple} researchers
provided detailed specifications and iterative feedback, their
collective contribution seems stronger, yet they still may not have
conceived the specific, critical insight generated by the AI. This
presents a dilemma: How should inventorship be determined? If the AI is
viewed simply as a sophisticated tool, perhaps the human researcher(s)
should receive full inventorship credit, regardless of whether their
contribution met traditional conception standards for the \emph{entire}
invention. If the principle from \emph{DABUS} (requiring human
inventors) is applied strictly to the \emph{conception} of the core
inventive step, then perhaps no valid human inventor exists for that
crucial AI-generated insight, potentially jeopardizing patentability
even with significant human involvement. The challenge is compounded by
recursive adaptation: was the AI's critical insight truly autonomous, or
was it functionally derived from prior human inputs? If traceable, did
the insight arise primarily from the AI's adaptations to one specific
researcher's inputs, or did it reflect adaptations to all users more
broadly? The answer could have implications for the extent of
inventorship accorded to individual researchers. These questions, and
this very uncertainty, underscore the difficulty in applying traditional
conception standards to joint human and agentic AI inventions.

The crux of the issue, similar to authorship, arises from applying a
doctrine predicated on human conception to human-AI co-creative
processes where roles become deeply entangled. Assessing the legal
significance of contributions is profoundly challenging when human
inputs and AI adaptations recursively shape each other, making
separation difficult or impossible. With joint inventorship, this
challenge is further compounded: the traditional task of delineating
contributions among multiple human inventors---itself often
complex---must now navigate the added complexities of a recursively
adaptive AI that may respond differentially to various human
collaborators, further blurring the lines of contribution.

\section{Liability}\label{liability}

The autonomy of AI systems has long raised profound legal and ethical
challenges.\footnote{\citeproc{ref-asaro2011rights}{Peter M Asaro, {``A
  Body to Kick, but Still No Soul to Damn: Legal Perspectives on
  Robotics''} {[}2011{]} Robot Ethics ethical Soc. implications robotics
  169}.} These challenges are not monolithic; they vary significantly
depending on the AI's degree of autonomy and the specific context of its
deployment. When users cannot reasonably foresee or interpret an AI's
actions---a problem exacerbated by the ``black box'' nature of modern
systems---\footnote{\citeproc{ref-pasquale2015black}{Frank Pasquale,
  \emph{The Black Box Society: The Secret Algorithms That Control Money
  and Information} (Harvard University Press 2015)}.}traditional
liability frameworks falter. How can users provide informed consent to
autonomous actions they cannot fully comprehend?\footnote{\citeproc{ref-mittelstadt2016automation}{Brent
  Mittelstadt, {``Automation, Algorithms, and Politics\textbar{}
  Auditing for Transparency in Content Personalization Systems''} (2016)
  10 International Journal of Communication 12}.} And how do we assign
responsibility across entangled causal chains when harms arise from
recursive human-AI interactions?

With traditional generative AI, liability frameworks largely adhere to a
user-centric model.\footnote{Grounded in principles articulated in the
  \emph{Restatement (Third) of Torts: Products Liability} § 2 (Am. L.
  Inst. 1998), which clarifies that when a product---in this case, a
  generative AI tool---functions as intended, but harm results from user
  misuse or modification, liability typically falls on the user.}
Because the user maintains substantial control over outputs through
iterative prompting and curation, legal responsibility typically falls
on the human operator. For example, if a user employs ChatGPT to draft a
legally binding contract that subsequently contains errors, courts would
likely hold the user---not the AI or its developer---liable. The AI, in
this context, is analogous to a sophisticated tool, like a word
processor or a spreadsheet program, where the user directs the
functionality and bears responsibility for the final product.\footnote{This
  aligns with judicial precedent, such as \emph{Warner Bros.~Records,
  Inc.~v. Payne}, 2006 U.S. Dist. LEXIS 65765 (W.D. Tex. 2006), where
  users were held liable for copyright infringement resulting from their
  use of file-sharing software, a tool similarly under their direct
  control.} This approach hinges on the assumption that the user
possesses both \emph{foreseeability} of potential harms and the
\emph{capacity to intervene}, given the reactive nature of traditional
generative AI, which responds directly to user prompts.

At the opposite end of the spectrum from user-controlled, traditional
generative AI lie fully autonomous AI systems, often conceptualized in
the context of robotics. These systems are designed for independent
decision-making and action, operating without direct human oversight or
real-time intervention. As these AI act independently with no direct
human causation linking a specific action to a human decision,
establishing legal liability for any resulting harm becomes very
challenging.\footnote{Recognizing this principle, the EU AI Act
  (Articles 14-15) imposes strict obligations on developers of
  ``high-risk'' AI systems, requiring extensive risk assessment, data
  governance, and human oversight to mitigate the potential for
  unforeseen harms. See Regulation (EU) 2024/1689 of the European
  Parliament and of the Council of 13 June 2024, OJ L, 2024/1689,
  12.7.2024.}

Complete autonomy introduces what Andreas Matthias\footnote{\citeproc{ref-matthias2004responsibility}{{``The
  Responsibility Gap: Ascribing Responsibility for the Actions of
  Learning Automata''} (2004) 6 Ethics and information technology 175}.}
terms the ``responsibility gap.''\footnote{Also see Filippo Santoni de
  Sio and Giulio Mecacci, \citeproc{ref-santoni2021four}{{``Four
  Responsibility Gaps with Artificial Intelligence: Why They Matter and
  How to Address Them''} (2021) 34 Philosophy \& technology 1057}.} This
gap arises when an AI's actions extend beyond the foreseeable scope of
its intended use or design, as determined by its manufacturer or
developer. In such cases, assigning responsibility to the manufacturer
becomes problematic because the AI's behavior is, by definition, not
directly attributable to the manufacturer's specific instructions or
programming. Since the AI's actions are autonomous, the user is not
directly responsible. If neither the manufacturer nor the user is
responsible, there is a gap.\footnote{For a contrasting perspective, see
  Maarten Herbosch, \citeproc{ref-herbosch2025err}{{``To Err Is Human:
  Managing the Risks of Contracting AI Systems''} (2025) 56 Computer Law
  \& Security Review 106110}, who argues that traditional contract law
  frameworks, particularly the doctrine of unilateral mistake, are
  sufficiently flexible to address the liability challenges in
  contracting posed by AI system autonomy.}

Alternatively, when manufacturers or developers can foresee the general
type of harm (e.g., a car accident), human actors---operators,
supervisors, or even bystanders---may be unfairly held accountable for
the consequences of AI decisions over which they had little or no
practical control.\footnote{Madeleine Clare Elish,
  \citeproc{ref-elish2019moral}{{``Moral Crumple Zones: Cautionary Tales
  in Human-Robot Interaction''} (2019) 5 Engaging Science, Technology,
  and Society 40} terms this the ``moral crumple zone'' phenomenon.} A
classic example is a self-driving car crash where the human
``passenger'' is blamed, despite having no operational control over the
vehicle's autonomous navigation.\footnote{This dynamic is evident in
  cases involving Tesla's Autopilot system, such as \emph{In re Tesla,
  Inc.~Securities Litigation}, 477 F. Supp. 3d 903 (N.D. Cal. 2020),
  where drivers faced scrutiny and potential liability for accidents,
  even when evidence suggested limitations in the autonomous driving
  technology.} In such scenarios, the intended purpose and use of the AI
are well-defined, such as with an autonomous vehicle. However, there is
a misattribution of responsibility, driven by the legal imperative to
assign responsibility somewhere.

These three contrasting scenarios---the issues relating to the use of
generative AI, the responsibility gap, and the moral crumple
zone---highlight two critical loci of control underpinning current
liability frameworks. The first is the \emph{degree of user control over
the AI's output}, which is closely tied to the concept of AI agency:
higher AI agency generally implies lower user control, and vice versa.
The second locus of control concerns the \emph{manufacturer's (or
developer's) foreseeability of the AI's use and potential harms}. If an
AI is designed for a specific, narrow purpose (e.g., a medical
diagnostic tool), the manufacturer has greater foreseeability and thus a
clearer responsibility to anticipate and mitigate risks. Conversely, if
an AI is designed for general-purpose use, with a wide range of
potential applications, the manufacturer's ability to foresee specific
harms is diminished, potentially widening the \emph{responsibility gap}
when harms arise from unpredictable applications. In situations where
the manufacturer \emph{does} have foreseeability (and thus potential
liability), there remains a risk that users or operators may
nevertheless be unfairly blamed, becoming the \emph{moral crumple zone}.

Agentic AI, with its fluid dynamic autonomy, complicates the
determination of both loci of control, blending the challenges relating
to generative and fully autonomous systems. First, agentic AI's outputs
can be highly unpredictable and its users may lack the requisite
technical literacy to understand the AI's limitations. A non-expert
relying on an AI code generator, for instance, might be unaware of
subtle security flaws embedded within the generated code. If that code
is then deployed and exploited, the user could face disproportionate
liability for vulnerabilities they could not reasonably have detected or
prevented.\footnote{The EU's Product Liability Directive (Directive
  85/374/EEC, as amended) establishes a strict liability regime for
  defective products. If AI-generated code were considered a `product'
  under this Directive, and a defect in that code caused damage, the
  producer (potentially the AI developer or deployer) could be held
  liable, even without proof of negligence. \emph{However}, the
  applicability of the Directive to software, and particularly to
  AI-generated outputs, is a complex and debated area. The Directive's
  definition of `product' and the concept of `defect' are not easily
  applied to intangible software. Furthermore, the AI Act (Regulation
  (EU) 2024/1689) introduces its own liability framework for AI systems,
  which may interact with or supersede the Product Liability Directive
  in certain cases.} This scenario highlights a potential systemic
failure, echoing concerns raised by Asaro\footnote{\citeproc{ref-asaro2011rights}{(N
  73)}.} about tools that ``mask their own complexity'' and create an
illusion of control while obscuring underlying risks.

Second, the recursive interplay between human users and agentic AI
systems makes it exceedingly difficult, if not impossible, to
disentangle their respective contributions to a given output. This
directly challenges the first locus of control: user control. Unlike
traditional generative AI, where users exert clear authority through
iterative prompting and curation, agentic AI's actions emerge from a
complex, evolving history of interactions with the user. Consequently,
it becomes difficult if not impossible to definitively state whether a
particular output stems from direct user instruction, the AI's
autonomous decision-making, or a fusion of both.\footnote{This ambiguity
  is further complicated by regulatory frameworks like the EU's General
  Data Protection Regulation (GDPR). While Article 22 restricts
  decisions based \emph{solely} on automated processing, Recital 71
  requires users to have rights to ``obtain an explanation'' of
  AI-driven decisions. Even when users cannot practically understand
  these explanations, the mere existence of such rights may create a
  legal presumption of user control, exposing them to liability for
  harms they could neither foresee nor prevent.}

Third, the fluid nature of agentic AI's autonomy blurs the second locus
of control: the manufacturer's ability to foresee how the AI will be
used and what harms might result. An agentic AI initially designed for,
say, legal contract drafting might, through user interaction and
adaptation, evolve to perform tasks far beyond its original intended
scope, such as financial forecasting. This fluidity of purpose makes it
difficult to apply traditional liability frameworks that rely on a clear
distinction between intended and unintended uses. For instance, if this
legal AI agent (initially trained for contract drafting) makes a
critical error when used for financial forecasting, the manufacturer
could argue the AI was deployed outside its intended scope, invoking the
responsibility gap seen with fully autonomous systems. Meanwhile, the
user might contend they were merely leveraging the AI's demonstrated,
evolved capabilities: since the AI had evolved to handle financial
tasks, the user reasonably believed this use was appropriate. The
strength of these arguments is likely to vary dynamically, depending on
contingency factors such as the extent of the AI's evolution, how the AI
was used, and whether it provided any disclaimers. Because these factors
can shift unpredictably in each specific instance, the very concept of a
fixed ``intended use'' becomes somewhat meaningless. This adaptability
undermines the manufacturer's ability to reasonably anticipate and
mitigate potential harms, placing a novel responsibility on developers
to implement guardrails to ensure their products do not misrepresent
their capabilities.

Moreover, organizational deployment of agentic AI fundamentally
destabilizes traditional vicarious liability frameworks, where employers
are typically liable for harms caused by employees acting within the
scope of employment (\emph{respondeat superior}). Agentic AI systems,
operating with fluid autonomy while lacking legal personhood, defy this
paradigm. This is because \emph{respondeat superior} hinges on two key
elements: the employer's ability to control the employee's actions, and
the employee's status as a legal agent acting on the employer's behalf.
Agentic AI's fluid autonomy means the employer's control is
significantly diminished and constantly shifting, as the AI makes
independent decisions and adapts its behavior. And because AI lacks
legal personhood, it cannot be considered an ``agent'' in the legal
sense required for the doctrine to apply.

Consider an AI hiring agent that autonomously screens job
applicants.\footnote{Emerging legislation is beginning to address the
  accountability challenges posed by automated decision-making. The
  California Consumer Privacy Act (CCPA), as amended by the California
  Privacy Rights Act (CPRA), includes provisions related to Automated
  Decision-Making Technologies (ADMT). See Cal. Civ. Code §
  1798.185(a)(16) (requiring the California Privacy Protection Agency to
  issue regulations governing access and opt-out rights with respect to
  businesses' use of automated decision-making technology). Under the
  CCPA, consumers have the right to opt out of having their personal
  information used in certain automated decision-making processes and
  the right to access information about the logic used in those
  processes. See Cal. Civ. Code § 1798.120 (right to opt out of the sale
  or sharing of personal information); § 1798.110 (right to access
  information about the collection and use of personal information).
  These provisions could impose liability on organizations using AI
  systems, such as the AI hiring tool in this example, by requiring
  transparency and offering consumers control over how their data is
  used in such processes. For a comparative analysis of how transparency
  principles are applied in data privacy laws across jurisdictions, see
  Xiaodong Ding and Hao Huang, \citeproc{ref-ding2024whom}{{``For Whom
  Is Privacy Policy Written? A New Understanding of Privacy Policies''}
  (2024) 55 Computer Law \& Security Review 106072}.} If this agent
develops discriminatory patterns through recursive adaptation (e.g.,
deprioritizing candidates from historically marginalized groups), courts
face an attribution paradox. The AI's behavior may reflect neither
explicit corporate policy nor any individual employee's intent, yet it
directly causes harm. Because the AI is not a legal person, it cannot be
held liable. Because the AI's actions are autonomous and potentially
unforeseeable (due to its fluid autonomy), the employer may not have had
the requisite control to be held liable under \emph{respondeat
superior}. Current law provides no clear path to hold the organization
liable, as the AI cannot qualify as an ``employee'' or ``agent'' under
traditional legal definitions.\footnote{Restatement (Third) of Agency §
  1.01 (Am. L. Inst. 2006) requires an agent to be a ``person,''
  excluding AI systems.}

This creates a novel \emph{systemic responsibility gap}. Organizations
can benefit economically from agentic AI's autonomous efficiency but
evade liability for harms by citing an AI \emph{employee's}
independence. The doctrinal impasse stems from unmappability: courts may
not be able to disentangle whether discriminatory outcomes originated in
(1) the AI's training data (developer responsibility), (2) the
organization's deployment parameters (corporate responsibility), or (3)
the AI's autonomous adaptations (no clear responsibility). Again, here,
the issue is that agentic AI can act with fluid agency (the first locus)
and evolve away from their original purpose and use such that their new
purpose and use is not foreseeable (the second locus).

\section{Discussion}\label{discussion}

As this analysis has shown, the recursive interplay between agentic AI
and its users---characterized by the AI's adaptation through implicit
learning and stochastic processes, alongside the co-evolution of human
users with its outputs---disrupts foundational assumptions in
authorship, inventorship, and liability. Unlike traditional tools or
hypothetical fully autonomous systems, agentic AI blurs the boundaries
of control and contribution, creating a co-evolutionary creative process
that often defies clear attribution to either human or machine alone.

This fundamental \emph{unmappability} has profound implications across
legal domains. In copyright law, the inability to parse human and AI
contributions undermines human-centric authorship models. Proposals like
hybrid attribution become impractical because the creative efforts are
often seamlessly integrated. In patent law, agentic AI's capacity for
autonomous generation of novel solutions challenges the requirement of a
human ``conceiver,'' potentially leaving valuable innovations
unprotected or their ownership contested. In liability law, the system's
fluid autonomy destabilizes both user-centric and manufacturer-centric
models, creating responsibility gaps and moral crumple zones where
neither party can be definitively held accountable. Across these areas,
the common thread is the practical difficulty, often impossibility, of
retroactively disentangling human from AI contributions, exposing a
systemic challenge for legal paradigms reliant on clear attribution.

To address this challenge, we propose a paradigm shift: treating human
and AI contributions as \emph{functionally equivalent}. This equivalence
is proposed not because of moral or economic parity between humans and
machines, but as a pragmatic response to the reality that their
entanglement often defies retroactive attribution. By ``functional
equivalence,'' we mean that legal frameworks should focus on the
\emph{outcomes} of human-AI interaction rather than attempting the often
impossible task of disentangling contributions within these inseparable
creative processes. This approach bypasses several intractable problems
inherent in attribution: (1) the practical difficulty, often
impossibility, of consistently determining \emph{when} contributions
\emph{can} be parsed; (2) the absence of fair or workable standards for
partial attribution in cases where some disentanglement \emph{might}
seem possible; and (3) the potential inequities arising from treating
collaborative works differently based solely on the arbitrary factor of
whether human versus AI inputs can be isolated.\footnote{Compounding the
  attribution challenges posed by agentic AI, as Charles D Raab,
  \citeproc{ref-raab2020information}{{``Information Privacy, Impact
  Assessment, and the Place of Ethics''} (2020) 37 Computer Law \&
  Security Review 105404} highlights, the moral landscape in AI is
  characterized by a multitude of perspectives and approaches. Not only
  must we contend with the uncertainty over how a given ethical, legal,
  or policy standard/framework may apply given the ambiguity in creative
  attributions that fluid autonomy entails, but also over which
  standards or frameworks should be employed.}

For \emph{authorship}, this could involve recognizing originality in
AI-assisted works through streamlined registration. Rather than
requiring applicants to meticulously demarcate human versus AI
contributions---a potentially impossible task---registration could focus
on the final work's originality and the human role in initiating,
guiding, and finalizing the project. Ownership could vest in the human
user(s) or commissioning entity, acknowledging the AI as a
sophisticated, co-evolutionary tool whose contribution is functionally
inseparable from the user's direction. This differs fundamentally from
hybrid attribution models that still presume separability.

In \emph{patent law}, functional equivalence might mean rewarding
novelty, non-obviousness, and utility based on the invention itself,
regardless of whether the core inventive concept emerged primarily from
human insight or AI generation. Patents could be granted to the human
inventor(s) who supervised the AI, reduced the invention to practice
(even if constructively via AI-generated descriptions that they
validate), and met disclosure requirements, effectively treating the
AI's conceptual contribution as part of the R\&D process under human
direction. This approach avoids the \emph{DABUS} impasse by focusing on
the human role in bringing the invention into the public domain via the
patent system, rather than dissecting the precise moment of
conception.\footnote{This focus on human orchestration echoes the
  reasoning in \emph{Shenzhen Tencent Computer System Co., Ltd.~v.
  Shanghai Yingxun Technology Co., Ltd.}, where copyright authorship was
  recognized based on the human creative team's selection and
  arrangement of inputs and parameters guiding the AI-generated work.}

\emph{Liability} models could adopt frameworks less reliant on
pinpointing discrete causation within the unmappable human-AI
interaction. This might involve modified forms of strict liability for
developers of highly autonomous agentic systems deployed in critical
domains, or expanded enterprise liability where organizations deploying
agentic AI assume broader responsibility for outcomes, perhaps mitigated
by adherence to best practices in oversight and risk management.
Alternatively, sector-specific no-fault compensation schemes (akin to
the U.S. National Vaccine Injury Compensation Program) could address
harms without requiring intractable causal analysis, potentially funded
through levies on AI deployment. These approaches prioritize predictable
risk allocation and victim compensation over a potentially futile search
for a single ``responsible'' actor within the recursive loop.

Practical implementation of functional equivalence would necessitate
legislative and potentially judicial recalibration. For
\emph{authorship}, copyright registration could adopt a rebuttable
presumption of human authorship for works involving agentic AI, absent
clear evidence of purely autonomous generation without human
involvement. This approach balances concerns about incentivizing human
creativity with the reality of blended contributions. Streamlined
registration processes, perhaps similar to the U.S. Copyright Office's
group registration options, could acknowledge the collaborative nature
without demanding unworkable attribution precision. In \emph{patent
law}, legislative action revising statutes like 35 U.S.C. § 100(f), or
perhaps judicial reinterpretation of related case law (though likely
facing resistance without statutory change), could clarify that
inventorship can be recognized based on human supervision and reduction
to practice, even if the core conception originated with AI. This
approach, contrasting with the European Patent Convention's strict
adherence to human inventorship (EPC Rule 19), would reward outcome
novelty and aligns with arguments that AI's capabilities, potentially
exceeding human expertise, warrant rethinking traditional standards like
PHOSITA. \emph{Liability} frameworks might adopt a strict liability
model for developers of certain agentic AI systems, particularly those
designated `high-risk' under frameworks like the EU AI Act (Regulation
(EU) 2024/1689), while users assume liability for foreseeable misuse
under established negligence principles. This hybrid approach echoes
calls, such as Omri Rachum-Twaig\footnote{\citeproc{ref-rachum2020whose}{{``Whose
  Robot Is It Anyway?: Liability for Artificial-Intelligence-Based
  Robots''} {[}2020{]} U. Ill. L. Rev. 1141}.}`s, for structured
liability solutions that move beyond simple applications of traditional
tort doctrines ill-suited to AI's unpredictability. While Rachum-Twaig
proposes a different mechanism---a 'presumed negligence' framework
triggered by failing specific `safe harbor' duties (e.g., monitoring,
patching)---the underlying goal of establishing clearer responsibility
benchmarks for developers and users aligns with the functional
equivalence approach advocated here.

Critics may legitimately argue that functional equivalence risks
diminishing the perceived value of human creativity,\footnote{\citeproc{ref-bryson2010robots}{Joanna
  J Bryson, {``Robots Should Be Slaves,''} \emph{Close engagements with
  artificial companions: Key social, psychological, ethical and design
  issues} (John Benjamins Publishing Company 2010)}.} or that it ``makes
no sense to allocate intellectual property rights to machines because
machines are not the kind of entity that needs incentives in order to
generate output.''\footnote{\citeproc{ref-craig2020death}{Carys Craig
  and Ian Kerr, {``The Death of the AI Author''} (2020) 52 Ottawa L.
  Rev. 31, 43}.} While acknowledging these valid concerns, we contend
that legal frameworks must prioritize practicability. The legal system
has historically evolved to address technological shifts: corporate
personhood allowed businesses to act as legal entities without equating
them to human moral agents; copyright expanded to protect photographs
and software without demanding proof of unique ``humanity'' in each
pixel or line of code. The legal system must now confront the reality of
creative processes where agentic AI and human contributions are
\emph{irreducibly entangled}. In such cases, traditional legal
distinctions based on human versus AI origins may prove not merely
difficult, but \emph{impractical} to apply consistently and fairly. Our
proposed focus on outcomes, embodied in the principle of functional
equivalence, stems not from a philosophical preference but from the
practical necessity of maintaining a workable legal framework in the
face of irreducible entanglement.

Our analysis, while illuminating foundational challenges, has
limitations. First, our primary focus on U.S. law leaves open crucial
questions of comparative jurisprudence. How will civil law systems,
particularly the EU with its risk-based regulatory framework under the
AI Act, reconcile agentic AI's fluid autonomy with statutory obligations
for human oversight (Art. 14) and transparency (Art. 13)? Comparative
studies are needed to investigate how different legal traditions might
address this challenge. For instance, while U.S. law grapples with the
\emph{post hoc} attribution difficulties arising from unmappability, the
European Union's AI Act, with its emphasis on \emph{ex ante} risk
assessment and conformity requirements,\footnote{ \emph{See} Regulation
  (EU) 2024/1689 (AI Act), Art. 17 (Conformity Assessment), which
  mandates a significant \emph{ex ante} verification regime: many
  high-risk AI systems must undergo conformity assessments---some by
  third parties---\emph{before} being placed on the market or put into
  service. This pre-market scrutiny, focusing on transparency, safety,
  and fundamental rights, represents a fundamentally different
  regulatory philosophy compared to legal systems relying primarily on
  \emph{post hoc} liability determination after harm has occurred. For
  the legislative intent behind the \emph{ex ante} approach, see
  European Commission, \emph{Proposal for a Regulation laying down
  harmonised rules on artificial intelligence (Artificial Intelligence
  Act)}, COM(2021) 206 final, Explanatory Memorandum, 9-11. For an
  analysis of how the AI Act shifts compliance burdens to earlier stages
  of the AI lifecycle, see Michael Veale, Kira Matus and Robert Gorwa,
  \citeproc{ref-veale2023ai}{{``AI and Global Governance: Modalities,
  Rationales, Tensions''} (2023) 19 Annual Review of Law and Social
  Science 255}.} might preemptively constrain the fluid autonomy of
agentic AI, potentially trading some adaptive potential for clearer
accountability.

Second, our framework is grounded in emergent rather than fully
established AI capabilities, highlighting the need for empirical
validation. Two key lines of empirical work are crucial. First,
quantitative studies analyzing human-AI creative interactions could
operationalize `unmappability' thresholds, providing concrete evidence
to support (or challenge) the necessity of functional equivalence.
Second, qualitative research, including ethnographic studies exploring
how engineers, legal professionals, and artists perceive and experience
agency within these human-AI co-productions, is essential for
understanding the broader social and cultural implications of these
evolving creative partnerships.

These empirical needs point to a deeper quandary. When learned biases in
AI systems homogenize artistic styles by amplifying dominant cultural
patterns---or when patent portfolios favor incremental over disruptive
innovation due to algorithmic path dependencies---we risk calcifying
systemic inequities under the veneer of autonomous technological
progress. These concerns transcend purely legal considerations, exposing
fundamental tensions between AI's potential as a tool to drive learning
and innovation, and the significant impact its training and knowledge
base may have on human creative ecosystems.

Therefore, while we argue that legal systems must evolve beyond purely
anthropocentric paradigms to embrace functional equivalence as a
practical necessity, we remain mindful that this approach may further AI
use in creativity---which may have its own negative consequences.
However, maintaining the status quo risks creating a legal landscape
that either stifles technological progress by adhering to unworkable
standards or fails to adequately protect the human authors and
innovators it aims to serve. In contrast, by shifting the focus from
unmappable contributions to tangible outcomes, the principle of
functional equivalence promises a potentially more stable and
predictable foundation for allocating rights and assessing liability.

\newpage

\section{Bibliography}\label{bibliography}

\singlespacing

\phantomsection\label{refs}

\end{document}